
%
\newbox\leftpage \newdimen\fullhsize \newdimen\hstitle \newdimen\hsbody
\tolerance=1000\hfuzz=2pt
\def\bigans{b }
\magnification=1200\baselineskip=12pt plus 2pt minus 1pt
\hsbody=\hsize \hstitle=\hsize 
%
%
\catcode`\@=11 
\newcount\yearltd\yearltd=\year\advance\yearltd by -1900

\def\Title#1#2{\nopagenumbers\abstractfont\hsize=\hstitle\rightline{#1}%
\vskip 1in\centerline{\titlefont #2}\abstractfont\vskip .5in\pageno=0}
\def\Date#1{\vfill\leftline{#1}\tenpoint\supereject\global\hsize=\hsbody%
\footline={\hss\tenrm\folio\hss}}
\def\draftmode{\def\draftdate{{\rm preliminary draft:
\number\month/\number\day/\number\yearltd\ \ \hourmin}}%
\headline={\hfil\draftdate}\writelabels\baselineskip=12pt plus 2pt minus 2pt
{\count255=\time\divide\count255 by 60 \xdef\hourmin{\number\count255}
	\multiply\count255 by-60\advance\count255 by\time
   \xdef\hourmin{\hourmin:\ifnum\count255<10 0\fi\the\count255}}}

\def\nolabels{\def\eqnlabel##1{}\def\eqlabel##1{}\def\reflabel##1{}}
\def\writelabels{\def\eqnlabel##1{%
{\escapechar=` \hfill\rlap{\hskip.09in\string##1}}}%
\def\eqlabel##1{{\escapechar=` \rlap{\hskip.09in\string##1}}}%
\def\reflabel##1{\noexpand\llap{\string\string\string##1\hskip.31in}}}
\nolabels
%
\global\newcount\secno \global\secno=0
\global\newcount\meqno \global\meqno=1
\def\newsec#1{\global\advance\secno by1
\xdef\secsym{\the\secno.}\global\meqno=1
\bigbreak\bigskip
\noindent{\bf\the\secno. #1}\par\nobreak\medskip\nobreak}
\xdef\secsym{}
\def\appendix#1#2{\global\meqno=1\xdef\secsym{\hbox{#1.}}\bigbreak\bigskip
\noindent{\bf Appendix #1. #2}\par\nobreak\medskip\nobreak}
%
%
\def\eqnn#1{\xdef #1{(\secsym\the\meqno)}%
\global\advance\meqno by1\eqnlabel#1}
\def\eqna#1{\xdef #1##1{\hbox{$(\secsym\the\meqno##1)$}}%
\global\advance\meqno by1\eqnlabel{#1$\{\}$}}
\def\eqn#1#2{\xdef #1{(\secsym\the\meqno)}\global\advance\meqno by1%
$$#2\eqno#1\eqlabel#1$$}
%
\newskip\footskip\footskip14pt plus 1pt minus 1pt 
\def\f@@t{\baselineskip\footskip\bgroup\aftergroup\@foot\let\next}
\setbox\strutbox=\hbox{\vrule height9.5pt depth4.5pt width0pt}
\global\newcount\ftno \global\ftno=0
\def\foot{\global\advance\ftno by1\footnote{$^{\the\ftno}$}}
%
%
\global\newcount\refno \global\refno=1
\newwrite\rfile
\def\ref{[\the\refno]\nref}
\def\nref#1{\xdef#1{[\the\refno]}\ifnum\refno=1\immediate
\openout\rfile=refs.tmp\fi\global\advance\refno by1\chardef\wfile=\rfile
\immediate\write\rfile{\noexpand\item{#1\ }\reflabel{#1}\pctsign}\findarg}
\def\findarg#1#{\begingroup\obeylines\newlinechar=`\^^M\pass@rg}
{\obeylines\gdef\pass@rg#1{\writ@line\relax #1^^M\hbox{}^^M}%
\gdef\writ@line#1^^M{\expandafter\toks0\expandafter{\striprel@x #1}%
\edef\next{\the\toks0}\ifx\next\em@rk\let\next=\endgroup\else\ifx\next\empty%
\else\immediate\write\wfile{\the\toks0}\fi\let\next=\writ@line\fi\next\relax}}
\def\striprel@x#1{} \def\em@rk{\hbox{}} {\catcode`\%=12\xdef\pctsign{
\def\semi{;\hfil\break}
\def\addref#1{\immediate\write\rfile{\noexpand\item{}#1}} 
\def\listrefs{\immediate\closeout\rfile
\baselineskip=12pt\centerline{{\bf References}}\bigskip{\frenchspacing%
\escapechar=` \input refs.tmp\vfill\eject}\nonfrenchspacing}
\def\startrefs#1{\immediate\openout\rfile=refs.tmp\refno=#1}
\def\figures{\centerline{{\bf Figure Captions}}\medskip\parindent=40pt}
\def\fig#1#2{\medskip\item{Figure ~#1:  }#2}
\catcode`\@=12 
%
\ifx\answ\bigans
\font\titlerm=cmr10 scaled\magstep3 \font\titlerms=cmr7 scaled\magstep3
\font\titlermss=cmr5 scaled\magstep3 \font\titlei=cmmi10 scaled\magstep3
\font\titleis=cmmi7 scaled\magstep3 \font\titleiss=cmmi5 scaled\magstep3
\font\titlesy=cmsy10 scaled\magstep3 \font\titlesys=cmsy7 scaled\magstep3
\font\titlesyss=cmsy5 scaled\magstep3 \font\titleit=cmti10 scaled\magstep3
\else
\font\titlerm=cmr10 scaled\magstep4 \font\titlerms=cmr7 scaled\magstep4
\font\titlermss=cmr5 scaled\magstep4 \font\titlei=cmmi10 scaled\magstep4
\font\titleis=cmmi7 scaled\magstep4 \font\titleiss=cmmi5 scaled\magstep4
\font\titlesy=cmsy10 scaled\magstep4 \font\titlesys=cmsy7 scaled\magstep4
\font\titlesyss=cmsy5 scaled\magstep4 \font\titleit=cmti10 scaled\magstep4
\font\absrm=cmr10 scaled\magstep1 \font\absrms=cmr7 scaled\magstep1
\font\absrmss=cmr5 scaled\magstep1 \font\absi=cmmi10 scaled\magstep1
\font\absis=cmmi7 scaled\magstep1 \font\absiss=cmmi5 scaled\magstep1
\font\abssy=cmsy10 scaled\magstep1 \font\abssys=cmsy7 scaled\magstep1
\font\abssyss=cmsy5 scaled\magstep1 \font\absbf=cmbx10 scaled\magstep1
\skewchar\absi='177 \skewchar\absis='177 \skewchar\absiss='177
\skewchar\abssy='60 \skewchar\abssys='60 \skewchar\abssyss='60
\fi
\skewchar\titlei='177 \skewchar\titleis='177 \skewchar\titleiss='177
\skewchar\titlesy='60 \skewchar\titlesys='60 \skewchar\titlesyss='60
\def\titlefont{\def\rm{\fam0\titlerm}
\textfont0=\titlerm \scriptfont0=\titlerms \scriptscriptfont0=\titlermss
\textfont1=\titlei \scriptfont1=\titleis \scriptscriptfont1=\titleiss
\textfont2=\titlesy \scriptfont2=\titlesys \scriptscriptfont2=\titlesyss
\textfont\itfam=\titleit \def\it{\fam\itfam\titleit} \rm}
\ifx\answ\bigans\def\abstractfont{\tenpoint}\else
\def\abstractfont{\def\rm{\fam0\absrm}
\textfont0=\absrm \scriptfont0=\absrms \scriptscriptfont0=\absrmss
\textfont1=\absi \scriptfont1=\absis \scriptscriptfont1=\absiss
\textfont2=\abssy \scriptfont2=\abssys \scriptscriptfont2=\abssyss
\textfont\itfam=\bigit \def\it{\fam\itfam\bigit}
\textfont\bffam=\absbf \def\bf{\fam\bffam\absbf} \rm} \fi
\def\tenpoint{\def\rm{\fam0\tenrm}
\textfont0=\tenrm \scriptfont0=\sevenrm \scriptscriptfont0=\fiverm
\textfont1=\teni  \scriptfont1=\seveni  \scriptscriptfont1=\fivei
\textfont2=\tensy \scriptfont2=\sevensy \scriptscriptfont2=\fivesy
\textfont\itfam=\tenit \def\it{\fam\itfam\tenit} 
\textfont\bffam=\tenbf \def\bf{\fam\bffam\tenbf} \rm}
%
%
\def\noblackbox{\overfullrule=0pt}
\hyphenation{anom-aly anom-alies coun-ter-term coun-ter-terms}
\def\inv{^{\raise.15ex\hbox{${\scriptscriptstyle -}$}\kern-.05em 1}}
\def\dup{^{\vphantom{1}}}
\def\Dsl{\,\raise.15ex\hbox{/}\mkern-13.5mu D} 
\def\dsl{\raise.15ex\hbox{/}\kern-.57em\partial}
\def\del{\partial}
\def\Psl{\dsl}
\def\tr{{\rm tr}} \def\Tr{{\rm Tr}}
\font\bigit=cmti10 scaled \magstep1
\def\biglie{\hbox{\bigit\$}} 
\def\lspace{\ifx\answ\bigans{}\else\qquad\fi}
\def\lbspace{\ifx\answ\bigans{}\else\hskip-.2in\fi} 
\def\boxeqn#1{\vcenter{\vbox{\hrule\hbox{\vrule\kern3pt\vbox{\kern3pt
	\hbox{${\displaystyle #1}$}\kern3pt}\kern3pt\vrule}\hrule}}}
\def\mbox#1#2{\vcenter{\hrule \hbox{\vrule height#2in
		\kern#1in \vrule} \hrule}}  
%
\def\CAG{{\cal A/\cal G}}   
\def\CA{{\cal A}} \def\CC{{\cal C}} \def\CF{{\cal F}} \def\CG{{\cal G}} 
\def\CL{{\cal L}} \def\CH{{\cal H}} \def\CI{{\cal I}} \def\CU{{\cal U}}
\def\CB{{\cal B}} \def\CR{{\cal R}} \def\CD{{\cal D}} \def\CT{{\cal T}}
\def\e#1{{\rm e}^{^{\textstyle#1}}}
\def\grad#1{\,\nabla\!_{{#1}}\,}
\def\gradgrad#1#2{\,\nabla\!_{{#1}}\nabla\!_{{#2}}\,}
\def\ph{\varphi}
\def\psibar{\overline\psi}
\def\om#1#2{\omega^{#1}{}_{#2}}
\def\vev#1{\langle #1 \rangle}
\def\lform{\hbox{$\sqcup$}\llap{\hbox{$\sqcap$}}}
\def\darr#1{\raise1.5ex\hbox{$\leftrightarrow$}\mkern-16.5mu #1}
\def\lie{\hbox{\it\$}} 
\def\ha{{1\over2}}
\def\half{{\textstyle{1\over2}}} 
\def\roughly#1{\raise.3ex\hbox{$#1$\kern-.75em\lower1ex\hbox{$\sim$}}}

\font\names=cmbx10 scaled\magstep0
\font\small=cmr10 scaled\magstep0

\def\i{{\rm i}}
\def\k{{\rm k}}
\def\eq{{\rm eq}}
\def\obj{{\rm obj}}
\def\Mpc{{\rm Mpc}}
\def\gal{{\rm g}}
\def\fin{{\rm f}}
\def\T{{\rm T}}
\def\S{{\rm S}}
\def\A{{\rm A}}
\def\sphere{{\rm sphere}}
\def\cube{{\rm cube}}
\def\obs{{\rm obs}}
\def\CDM{{\rm CDM}}
\def\sg{{\sigma_8({\rm gal})}}
\def\sc{{\sigma_8({\rm CDM})}}
\def\bias{{\sigma_8^{-1}({\rm CDM})}}
\def\met{g^{\mu\nu}}
\def\pm{\partial_\mu}
\def\pn{\partial_\nu}
\def\st{\theta(t)}
\def\sd{\dot \st}
\def\bps{{\bf\Phi} ({\bf\vec x}, t)}
\def\bp{\bf\Phi}

\baselineskip=20pt
\Title{
PUPT-1674}
{\vbox{\centerline { Inflationary Textures }}}
   \footnote{}{\small Email address:  $^\ddagger$N.G.Turok@damtp.cam.ac.uk,
 $^\dagger$zhu@puhep1.princeton.edu.}
\font\large=cmr10 scaled\magstep3
\centerline { \names Neil Turok$^{\dagger,\ddagger}$ and
        Yong Zhu$^\dagger$}
\centerline{\small  $^{\dagger}$Physics Department}
\centerline{\small  Princeton University, Princeton NJ 08544,}
\centerline{\small  $^{\ddagger}$DAMTP}
\centerline{ \small  Silver Street, Cambridge, U.K.}

\bigskip
\bigskip
\bigskip
\centerline{\bf Abstract}
\baselineskip=18pt

%
{\small We present a calculation of the power spectrum 
generated in a classically symmetry-breaking 
$O(N)$ scalar field through inflationary 
quantum fluctuations, using the 
large-$N$ limit.
The effective potential of the theory
in de Sitter space is obtained from a gap equation which is 
exact at large $N$.
Quantum fluctuations restore the $O(N)$ symmetry in de Sitter
space, but for the finite values of $N$ of interest,
there is symmetry breaking and phase ordering after inflation,
described by the classical nonlinear sigma model. 
The main difference with the usual cosmic texture scenario is that
the initial conditions possess long range correlations,
with calculable exponents.  }
 
\Date{\small 12/96} 

\baselineskip=18pt

\bigskip

\centerline{\bf 1. Introduction}

There are currently two broad categories of theories of
cosmological perturbation: those which generate structure
via quantum fluctuations during inflation
\ref\inf{
A.A. Starobinsky, Phys. Lett. {\bf 91B}, 99 (1980);
 A.H.  Guth, Phys. Rev.  {\bf D23}, 347 (1981); 
A.D. Linde, Phys. Lett. {\bf 108B}, 389 (1982); 
A. Albrecht and P. Steinhardt,
 Phys. Rev. Lett. {\bf 48}, 1220 (1982).}, 
and those which generate structure  upon the process of symmetry breaking and 
field ordering through the Kibble mechanism  
\ref\ki{T.W.B. Kibble, J. Phys.  {\bf A9}, 1387 (1976).}.
Inflation is a brief  phase of superluminal expansion in the early universe, 
which is invoked to solve the horizon problem, the flatness problem and other 
puzzles in the standard cosmology. 
The existence of an event horizon gives
rise to quantum fluctuations in the scalar field, which 
can be attributed to the Hawking temperature of de Sitter space.
These fluctuations are then stretched out to 
cosmological scales.
 As the wavelength of a given mode  
expands beyond the Hubble radius 
 $H^{-1}$, its amplitude becomes frozen.
A broad-band spectrum of fluctuations with this amplitude is generated
over the course of the inflation. Once inflation is over, the Hubble radius
grows faster than the scale factor $a(t)$. The
fluctuations 
stretched out of the the Hubble radius
 during inflation re-enter  after inflation with 
nearly same amplitude, giving an almost scale-invariant 
spectrum of perturbations 
 \ref\bst{S. W. Hawking, Phys. Lett. {\bf 115B}, 295 (1982);
A.A. Starobinsky, Phys. Lett. {\bf 117B}, 175 (1982); A.H. Guth and S.Y. Pi,
Phys. Rev. Lett. {\bf 49}, 1110 (1982); 
J. Bardeen, P. Steinhardt and M. Turner, Phys. Rev. {\bf 
D28}, 679 (1983).},\ref\kol{L.A. Kofman and A.D. Linde, Nucl. Phys. 
{\bf B282}, 555 (1986).}. 
Requiring that the density perturbations are
consistent with  the observed level of structure in today's universe
requires some fine-tuning of parameters in most inflationary models.

The symmetry breaking  category  
\ki,
includes the theories of 
cosmic strings\ref\vsbook{ For a review see A. Vilenkin and 
E.P.S. Shellard, ``{\it Cosmic Strings and other Topological Defects}'',
Cambridge University Press, 1994.},
global monopoles and textures  
\ref\texture{
N. Turok, Phys. Rev. Lett. {\bf 63}, 2625 (1989).}.  
A simple class of theories with broken global symmetry is provided by the
``$O(N)$'' models, where a global $O(N)$ symmetry group is spontaneously 
broken to $O(N-1)$ by a $N$-component scalar field ${\bf \vec\phi}$. 
For $N=1$, one has domain walls,
$N=2$, global strings and  $N=3$, global monopoles. For $N=4$, there are 
textures, and for $N>4$, ``nontopological textures''. In the limit of large
$N$, there are no topological defects, and the theory is exactly 
solvable \ref\larN{N. Turok and D. Spergel, Phys. Rev. Lett. {\bf
66}, 3093 (1991).}. In these pictures, 
the universe is usually 
assumed to begin in a hot, smooth and  homogeneous  state, perhaps created by
inflation or just special initial conditions.
As the universe  cools, the symmetry is broken and a disordered phase
forms, containing defects of the appropriate type.
As the universe expands, the tangle of defects is ordered
on progressively larger scales.
Density perturbations  of 
constant amplitude at horizon crossing are
generated in the process, 
resulting in a scale invariant power spectrum of perturbations.
The phase ordering process involves nonlinear physics, which leads to a 
nongaussian pattern of fluctuation \ref\nongau{D. Coulson, P. Ferreira, 
P. Graham, and N. Turok, Nature {\bf 368}, 27 (1994).},
\ref\ntnew{N. Turok, DAMTP preprint, bulletin board astro-ph/9606087 (1996).}
through 
which they may be distinguished from inflation.

At the current stage of a rapidly developing field, none of the simplest
versions of the  
above theories
provides a compelling match to 
the observations. Several variations of the simplest inflationary model
have been introduced in an attempt to improve matters:
tilting the power spectrum, introducing gravity waves, 
adding  hot dark matter or a cosmological constant, or 
considering open inflation. 
It is also surely worthwhile to consider other options.
One of the simplest is  
just to combine the topological defect  and
inflation pictures. One possibility is to have standard inflation, 
with defects produced after reheating in the usual manner. 
Such a theory would have inflationary and defect-induced fluctuations, 
in some linear combination, uncorrelated with each other. 
A detailed study of reheating after inflation is recently
been introduced by Kofman, Linde, and Starobinsky
 \ref\kls{L.A. Kofman, A.D. Linde, and A.A. Starobinsky, 
Phys. Rev. Lett. {\bf 76}, 1011 (1996).}, 
where parametric resonance may rapidly 
transfer most of the energy of an inflaton field to the energy
of other bosons. This process
may result in copious production of topological
defects. 
Perhaps more interesting is the case where
exponentially large defects are generated through phase
transitions during the last stages of inflation \kol.
 Vishniac, Olive and Seckel  \ref\vos{E.T. Vishniac, K.A. Olive,
and D.Seckel, Nucl. Phys. {\bf B289}, 717 (1987).} have proposed
 a class of models in which 
the inflaton is coupled to the string-producing field. And the strings are 
formed late in inflation as the inflaton rolls towards its zero-temperature
value. The case that cosmic strings and textures are produced  in 
inflation via  
quantum creation or sufficiently high reheat temperature has been considered 
by Hodges and Primack \ref\hp{H.M. Hodges, and J.R. Primack, Phys. Rev.
{\bf D43}, 3155 (1991).}. A second-order phase transition during inflation is 
studied as a natural mechanism to produce topological defects by 
Nagasawa snd Yokoyama \ref\ny{M. Nagasawa and J. Yokoyama, Nucl. Phys. 
{\bf B370}, 472 (1992).}.
Finally, 
quantum creation of topological defects through instanton tunneling during
inflation has been investigated by 
 Basu, Guth, and  Vilenkin   \ref\topinf{R. Basu, A.H. Guth, and A. Vilenkin,
Phy. Rev. {\bf D44}, 340 (1991).}. 

In this work, we aim at an analytic treatment of this problem using
large $N$ techniques, developing earlier work by R. Davis and one of
us
\ref\dt{R.L. Davis and
N. Turok, unpublished.}. We study a renormalisable symmetry beaking 
 $O(N)$ scalar field theory 
in the large
$N$ limit. We study the phenomenon of symmetry restoration in de Sitter space
\ref\ratra{B. Ratra, Phys. Rev {\bf D31}, 1931 (1985).}, 
and match the field modes
to those in the subsequent radiation dominated era. 
This provides a set of initial conditions for the usual sigma model evolution, 
but with  
long range correlations. 
The primordial scalar field 
power spectrum  $P_k$ is obtained as a function of the scalar 
self-coupling.

\vskip 10pt
\centerline{\bf 2. $\lambda \phi^4$ Theory in de Sitter Space}
We study the $\lambda \phi^4$ theory in four
dimensions with the lagrangian density
\eqn\lagr{ \eqalign{
L=\sqrt{-g}\biggl[\half \met\pm\phi^\alpha\pn\phi^\alpha-
\half (m_0^2 +\xi_0 R^2)|\vec\phi|^2 
- {\lambda_0
 \over 4! N}|\vec\phi|^4 \biggr] ~.}} 

\vskip 10pt
\centerline{\bf 2.1 Large $N$ Approximation}
The $1/N$ expansion of the effective potential is constructed  by 
summing  the ``super-daisy diagrams'' 
\ref\dj{L.  Dolan and R. Jackiw, 
Phys. Rev. {\bf D9}, 3320 (1974).}
in  the limit of $N \to \infty$ with fixed coupling
constant $\lambda $.  These 
formulas typically display richer structures than the corresponding
leading-order expressions in ordinary perturbation theory because
that the leading $1/N$ approximation preserves much more of the 
nonlinear structure of the exact theory than does ordinary 
lowest-order  perturbation theory. Large $N$ behavior of a massless
scalar field in flat spacetime was first studied by Coleman, 
Jackiw  and Politzer \ref\flatN{S. Coleman, R. Jackiw  and H.D. Politzer, 
Phys. Rev. {\bf D10}, 2491 (1975).}. 
As well known, the one-loop radiative correction
to the effective theory dynamically breaks the $O(N)$ symmetry 
\ref\cw{S. Coleman and 
E. Weinberg, Phys. Rev. {\bf D7}, 1888 (1973).}.  However, a tachyonic
excitation was  found in the 
symmetry breaking state at large $N,$  which cast some doubts
on the validity of the large $N$ approximation.
 Later,  Abbott, Kang  and Schnitzer \ref\flatn{L.F. Abbott, 
J.S. Kang  and J. Schnitzer,  Phys.
Rev. {\bf D13}, 2212 (1976).} and Linde \ref\liN{A.D. Linde, Nucl. 
Phys. {\bf B125}, 369 (1977).}
pointed out that
the effective potential of the scalar field $\phi$ is a double-valued 
function, with  one branch containing   the
symmetry breaking phase and the other including  the
symmetric solution. It was shown that the symmetric  state
{\it always} had  lower effective energy. Thus the ground state
of the theory is  $O(N)$-symmetric, and the spontaneous symmetry breaking
is not possible in the large-$N$ limit.
 No tachyon exists in the symmetric vacuum,
and the excited states can be studied through perturbative methods.

This poses a problem for us - in the large $N$ limit, the theory we
are discussing does not have a symmetry breaking phase,  even in
Minkowski space. So there would be no classical phase ordering
following inflation. However, we shall take the attitude that we
are not in any case really interested in very large $N$, (not
least because the amplitude of density perturbations goes to zero
as $1/\sqrt{N}$), but we merely wish to use the large $N$ approximation
as a tool to approximate a theory with $N=3,4$ or $5$. We know that
at these modest values of $N$ the approximation
{\it fails} in Minkowski space, so there is symmetry breaking.
It is nevertheless reasonable to hope that the approximation
{\it is} valid in de Sitter space, where one expects symmetry
restoration to occur for any $N>1$ \ratra. We shall therefore treat the
problem in the quantum mechanical large $N$ approximation in
de Sitter spaces, but the classical large $N$ approximation (which does
have symmetry breaking) in
the subsequent radiation era. This may seem a schizophrenic
point of view, but it is one we are forced to in order to have a chance of
reproducing even the qualitative physics correctly.

\vskip 10pt
\centerline{\bf 2.2 de Sitter Ground State}

We work in the flat slicing of  de Sitter spacetime 
with  conformal coordinate $\eta$ and metric 
\eqn\dw{ \eqalign{
d\tau^2 = a^2(\eta)[d\eta^2-(d{\bf x})^2] ~. }}
where the scale factor takes $a(t)=e^{Ht}$, or
\eqn\dw{ \eqalign{
a(\eta)=-{1\over H \eta}, \;\;\;\; \eta<0  ~. }}
We rescale $a(\eta)$ to be unity
in the beginning of inflationary epoch.
Thus at the  end of inflation, or the transition to the radiation 
dominated era,
we have $\eta_0=-\eta_{transition}\ll1/H$.

In curved spacetime, the massive  scalar  field with mass $m$ 
can be  expanded as
\eqn\dw{ \eqalign{
\phi^\alpha({\bf x},\eta)
       =(2\pi)^{-3/2} \int d^3{k}[\phi_k(\eta)\hat a^\alpha_{\bf k}e^
{i{\bf k\cdot x}}+\phi_k^*(\eta)
       \hat a^{\alpha\dagger}_{\bf k}e^{-i{\bf k\cdot x}}] ~,}}
where
\eqn\shit{ \eqalign{
       [{\hat a^a_{\bf k},\hat a^{b\dagger}_{\bf q}}] =
(2 \pi)^3\delta^{ab}\delta^3({\bf k-q})  ~.}}
The mode function $\phi_{_k}(\eta)$ can be written in a separated form as
 \ref\bdd{N.D. Birrell and P.C.W. Davies, ``{\it Quantum Fields in 
Curved Space}'', Cambridge University Press, 1982.}
$ 
\phi_{_k}(\eta)=a^{(2-n)/2}\chi_{_k}(\eta)~ .
$
The canonical quantization conditions
reduce
to a condition on the Wronskian of the solutions $\chi_{_k}$: 
$
\chi_{_k}\partial_\eta\chi_{_k}^\ast-\chi_{_k}^\ast\partial_\eta\chi_{_k}=i.
$
As usual, we user the Bunch-Davies vacuum \ref\bd{T.S. Bunch and P.C.W. Davies, Proc. Roy. 
Soc. (London) {\bf A360}, 117 (1978).}
for the  scalar field:
\eqn\eg{ \eqalign{
\phi_k(\eta)&=\sqrt{\pi\over4}H\eta^{3/2}H_\nu^{(2)}(k\eta) ~, \cr
\nu^2&={(n-1)^2\over4}-{ m^2\over H^2} ~, }}
where  $n$ is the spacetime dimension.  The two-point correlation
function can be expressed in dimensional regularization 
 \ref\dimreg{P. Candelas and D.J. Raine, Phys. Rev. {\bf D12}, 965 (1975).}
\eqn\pppi{ \eqalign{
G(x,x) = {H^2\over8\pi^2}
       {\Gamma(\nu(n)-{1\over2}+n/2)\Gamma(-\nu(n)-{1\over2}+n/2)
       \over\Gamma({1\over2}+\nu(n))\Gamma({1\over2}-\nu(n))}
       \Gamma(1-n/2) ~. }}

\vskip 10pt
\centerline{\bf 2.3 $1/N$ Approximation in de Sitter Space}

The summation of superdaisy diagrams in flat space time 
has been discussed extensively 
\ref\jac{ R. Jackiw, Phys. Rev. {\bf D9}, 1686 (1974).}. The similar 
calculation is carried out here in de Sitter metric.
We merely state the gap equation here, which follows from identical diagrammatics
to the flat spacetime case
\eqn\gapp{ \eqalign{
2 {d\,V_{eff}(\phi^2)\over d\phi^2}\equiv
 M_{ds}^2(\phi^2) 
 &= m_0^2 +\xi_0 R+
{\lambda_0 \over 6N}\phi^2+{\lambda_0 \over 6} \langle 
\phi^2 \rangle |_{M_{ds}} ~,}}
where  
$M_{ds}^2$ is  the effective mass squared of the theory in de Sitter
space.  Note that in the gap equation   \gapp~ 
$\phi$ is the classical field, while
 $\langle\phi^2\rangle$ is the two-point function 
given by equation \pppi.
From equation \pppi, the two-point function diverges  at 
$n=4.$
Its  renormalization is carried out in the 
standard procedure through 
dimensional regularization and minimal subtraction \dimreg:
\eqn\dw{ \eqalign{
m^2 &= m_0^2+{\lambda_0 \over 6} {|m|^{d-2} \over 
(4\pi)^{d/2}}\Gamma(1-d/2) ~,\cr
{1\over\lambda} &= {1\over\lambda_0}+{1\over 6}{|m|^{d-4} \over 
(4\pi)^{d/2}}\Gamma(2-d/2) ~,\cr
\xi &= \xi_0 +{\lambda_0 \over 6}d(d-1)({1\over 6}-\xi)
{|m|^{d-4} \over (4\pi)^{d/2}}\Gamma(2-d/2)  ~.\cr }}
We shall only study the simplest case: zero conformal coupling, since
the extension to nonzero $\xi$ is straightforward.

An alternative  approach to remove the divergence in the two-point
function is to apply an effective cutoff on the comoving 
momentum due to  the inflationary  
expansion \ref\renorm{A. Vilenkin, Nucl. Phys. {\bf B226}, 504 (1983);
{\it ibid}, 527 (1983).}. We shall take 
this approach to  verify the 
consistency  and validity of renormalization scheme in our model.

After renormalisation, the gap equation \gapp~ reads
\eqn\gap{ \eqalign{
M_{ds}^2 = m^2+ {16 \pi^2 \over N}g \phi^2+{1\over 2}g M_{ds}^2ln{M_{ds}^2
\over {|m^2|}}-12gH^2ln{M_{ds}^2\over {|m^2|}}
+g H^2\Biggl\{({M_{ds}^2\over H^2}-2) \cr
\biggl[ \Psi({3\over 2}+\nu)+
\Psi({3\over 2}-\nu) - ln{M_{ds}^2\over H^2}-1 \biggr]+{M_{ds}^2\over 
H^2}-{2\over 3}\Biggr\} ~,}}
where
\eqn\dw{ \eqalign{
g ={\lambda \over {96 \pi^2+\lambda}},\,\,\,\,  \nu^2={9 \over 4}- {M_{ds}^2 
\over H^2} ~ .}}
The effective potential can be computed by integrating $M_{ds}^2$
\eqn\dw{ \eqalign{
V(\Phi) &= \int\displaylimits_0^{\Phi} {dV\over d\phi}d\phi \;+ \;V(0)\cr
	      &= \int\displaylimits_0^{\Phi} \phi \cdot M_{ds}^2(\phi^2) \cdot d\phi \;+\; V(0)  .\cr  }}

\bigskip

\vskip 10pt
\centerline{\bf 3. FRW Solution and Mode Matching}

In the radiation dominated era, the field ordering 
is  described classically by the nonlinear sigma model \larN~.
The evolution of the N-component scalar field
$\vec \phi$   obeys 
\eqn\dw{ \eqalign{
{\partial}_\eta^2 \phi^\alpha +2 {\partial_\eta a(\eta) \over
a(\eta)}\partial_\eta \phi^\alpha -{\vec \nabla}^2 \phi^\alpha
=-{{(\partial {\vec \phi})^2}\over 
\phi_0^2}\phi^{\alpha} \equiv {T_0 \over \eta^2}\phi^\alpha .}}
In  the  large $N$ approximation,
 $T_0$ may be treated as a constant
once scaling behaviour is attained for the field ordering.
In momentum space, the decoupled mode functions satisfy
\eqn\rad{ \eqalign{
{d^2\phi_k\over {d(k\eta)^2}}+{\alpha \over {k\eta}}{d\phi_k\over {d(k\eta)}}+
(1-{T_0\over {(k\eta)^2}})\phi_k=0, }}
where $\alpha=2{dln(a(\eta))\over dln\eta}$ with  $\alpha=2$ for 
radiation-dominated era and $\alpha=4$ for  matter-dominated era. 
Note that equation \rad~  is symmetric under  $\eta \to -\eta$.
In the radiation dominated era, 
the solution of \rad ~can be generally expressed as 
\eqn\frwc{ \eqalign{
\phi_k &= \sqrt{\pi\over 4} H (-\eta)^{-{1\over 2}} 
\biggl [ C_1(k)H_\mu^{(1)}(-k\eta)+C_2(k)H_\mu^{(2)}(-k\eta)\biggr ], \cr
\mu^2&=T_0+{1 \over 4 }~.}}

After the end of inflation, the mode functions  \eg~ of $\phi$ in de Sitter
space   is matched to the FRW classical solution \frwc~
by requiring that scale factor $a(\eta),$
 $\phi_k(\eta),$ and  $\dot\phi_k(\eta)$
be continuous at the transition point $\eta=-\eta_0\;.$
A simple model of 
the transition can  be constructed by considering a metric of the form
\eqn\dw{ \eqalign{
a(\eta)\;=\;\cases{ \;-( H\eta)^{-1},\;\>\>\>\;\;\;\;
\;\;\;\;\;
\>\> \>\>\>\;\eta<-\eta_0,\>\>\>\;\; {\rm de ~Sitter }\cr
\;(\eta+2\eta_0)/ H\eta_0^2,\;\;\;\;\;\;\;\;\;\eta>-\eta_0,
\;\;\;\;{\rm FRW}}}}
and the matching condition reads
\eqn\matc{ \eqalign{
{\eta_0}^{2} H_\nu^{(2)}(k\eta_0) =  &\biggl 
[ C_1 H_\mu^{(1)}(k\eta_0) + C_2 H_\mu^{(2)}(k\eta_0)\biggr ]~ , \cr
\eta_0^2 {\dot H_\nu}^{(2)}(k\eta_0)
 + {3\over 2} \eta_0
H_\nu^{(2)}(k\eta_0) = &\biggl [C_1{\dot H_\mu}^
{(1)}(k\eta_0)+C_2 {\dot H_\mu}^{(2)}(k\eta_0) \biggr] \cr
-{1\over 2\;\eta_0} &\biggl 
[C_1 H_\mu^{(1)}(k\eta_0)+C_2 H_\mu^{(2)}(k\eta_0)\biggr ] ~,}}
where dot denotes $
\partial/\partial \eta_0$. The solution of \matc ~ is 
expressed as 
\eqn\dw{ \eqalign{
C_1 &= -{i\pi\,\eta_0\,\over 4}\biggl[ \eta_0^2 H_\nu^{(2)}(k\eta_0)B-
H_\mu^{(2)}(k\eta_0) C\biggr] ~,\cr
C_2 &=  {i\pi\,\eta_0\,\over 4}\biggl[ \eta_0^2
H_\nu^{(2)}(k\eta_0)A- H_\mu^{(1)}(k\eta_0) C\biggr] ~,}}
where the coefficients read
\eqn\dw{ \eqalign{
A &= {\dot H_\mu}^{(1)}(k\eta_0) - {1\over 2\eta_0} H_\mu^{(1)}(k\eta_0)~,\cr
B &= {\dot H_\mu}^{(2)}(k\eta_0) - {1\over 2\eta_0} H_\mu^{(2)}(k\eta_0)~,\cr
C &=\eta_0^2 {\dot H_\nu}^{(2)}(k\eta_0) + {3\over 2}\eta_0 
H_\nu^{(2)}(k\eta_0)~ .
}}
Easily we  can verify 
the normalization condition for Bunch-Davis vacuum. 
The normalization conditions
for the FRW mode functions are also automatically satisfied.

The mode function \frwc~  in FRW can be separated into a
`classical' part (by which we mean the component which survives
with large amplitude long after the transition to 
 radiation domination)
and a `quantum' part (which diminishes quickly after inflation):
\eqn\yong{ \eqalign{
\phi_k(\eta)\; &\equiv\;\phi_{cl}\;+\;\phi_q ~,\cr
&=\;a(k)\;J_\mu(k\eta)\;+\;b(k)\;Y_\mu(k\eta)\cr  ~,}}
where the coefficients $a$ and $b$ write
\eqn\coeff{ \eqalign{
a(k)\,&=\,H {\pi^{3\over 2} \, \eta_0\over 8}\biggl
 [\eta_0^2 \,H_\nu^{(2)}(k\eta_0)\;(2\dot Y_\mu(k\eta_0)
		-{Y_\mu(k\eta_0)\over \eta_0})\;-\;2\,C\,
Y_\mu(k\eta_0)\biggr ] ~,\cr
b(k)\,&=- H{ \pi^{3 \over 2} \,\eta_0\over 8}\biggl [\eta_0^2\, H_\nu^{(2)}(k\eta_0)
\;(2\dot J_\mu(k\eta_0)
	-{J_\mu(k\eta_0)\over \eta_0})\;-\;2\,C\,J_\mu(k\eta_0)\biggr ]~.
\cr}}
At the epoch of most importance for structure
formation, or when a given mode re-enters the Hubble radius,  $k\eta\approx1$,
the `classical' part dominates the `quantum'
part by a factor of $\Bigl| {\eta \over  \eta_0 }\Bigr|^{2\mu}$ with $\mu>0$.
Consider how small $\eta_0$ is, this is an extremely
large number. It is therefore reasonable to neglect the quantum part
of the wavefunction  at FRW space. 
Therefore, the classical non-linear sigma model is  a valid
approximation to the $O(N)$ scalar field at FRW epoch.

After matching  the quantum fluctuations of the scalar field to
the classical evolution  of the nonlinear sigma model, 
we can now relate the original parameters in 
the quantum field theory  to the value of 
$\vec \phi^2 = \phi_0^2$ 
in the classical theory. 
We have
\eqn\mmat{ \eqalign{ \mu=2-\nu ~.}} 
And at large $\eta \gg \eta_0$ 
\eqn\ddww{ \eqalign{
\phi_0^2\,\equiv\,N\langle \phi_{cl}^2 \rangle \,&=\,
N \int{{d^3k \over(2\pi)^3}{1\over \eta}\phi_{cl}^2} ~,\cr
&=N {H^2\over 4 \pi^4} {\Gamma^2(\nu) \Gamma^2(\mu)
 \Gamma(\mu-1/2)\over\Gamma(\mu+3/2)} ~.
}} %
At later time in FRW, the classical dynamics of the
$\lambda \phi^4$ theory
is totally fixed by the tree  diagrams. Thus, the renormalised values
$m^2$ and $\lambda$ are  fixed by
\eqn\emla{ \eqalign{
m^2= \;- {\lambda\over 6N} \phi_0^2(H)~,
}}
where $\ \phi_0^2(H)$ is a function of $H$ as fixed by \ddww.
To re-emphasize the point, we are treating the theory
as classical in the post-inflation era, with the given renormalised parameters,
with the justification that for modest values of $N$ and $\lambda$
this should be a reasonable approximation.
\bigskip

\vskip 10pt
\centerline{\bf 4. Effective Potential and Vacuum  in 
de Sitter Spacetime}
The effective potential of the $\lambda\phi^4$ theory in de Sitter space
is studied under the assumption of  equation \emla.
 After plugging  \emla~into the gap equation \gap,  our 
theory will have only one free parameter --- 
the renormalised coupling constant  $\lambda$.
The effective potential $V_{eff}$ is obtained  for 
different regions of $\lambda$ and the corresponding vacuum states are  
found  by minimizing  $V_{eff}$
\eqn\dv{ \eqalign{
0={dV_{eff}(\phi^2) \over d\phi} &\equiv 2\phi {dV_{eff}\over  d\phi^2}\cr
&=\phi \;M_{ds}^2(\phi^2) ~.\cr}}
To satisfy   equation \dv, the vacuum state will either have  $|\phi|=0$, where
the symmetry is restored  or
$M_{ds}^2=0$, where the symmetry is broken. 

\vskip 10pt
\centerline{\bf 4.1 Weak coupling:$\;\;\lambda\ll N$ }
Weak coupling implies the small mass limit: $ |M_{ds}^2|\ll H^2$.  
At first, let's study  the weak interacting  
limit  of the gap equation \gap $\,$ with $\lambda\ll N ~{\rm and}~ 
 |M_{ds}^2| \ll H^2$. The leading order terms in this limit
\eqn\vv{ \eqalign{
\nu\;&=\;\sqrt{{9\over 4}-{M_{ds}^2\over H^2}}\;=\;{3\over2}-{1\over3}
{M_{ds}^2\over H^2}+{\rm O}({M_{ds}^2\over H^2}) ~,\cr
\phi^2_0 &=  {3 N H^2\over 16 \pi^2}{H^2 \over M_{ds}^2} ~,\cr
m^2 &=-{\lambda \over 6N} \phi_0^2=-{\lambda H^2 \over 32 \pi^2}\cdot
{H^2\over M_{ds}^2} ~,\cr
\psi({3\over2}-\nu)\;&=\;\psi({M_{ds}^2\over 3 H^2})
\;= \;-{3 \over 2} { H^2\over M_{ds}^2}  ~. }}
The leading order of the gap equation \gap~is
\eqn\vv{ \eqalign{
{M_{ds}^2\over H^2} = \sqrt {\lambda \over 32 \pi^2}~ .}}
The variable $\nu(\lambda)$ can be expressed as 
$$
\nu={3\over2}-{1\over 12 \pi}\sqrt{{\lambda\over2}} ~.
$$
Generally, the relation of  $M_{ds}^2(\phi^2)$  over $\phi^2$ is
plotted at Fig.(1).
At the region  $M_{ds}^2<A$, $\phi^2$ is negative, 
which is obviously unphysical. 
Thus, $M_{ds}^2$ cannot be zero and   
the ground state of the scalar field
is the $O(N)$-symmetric state
\eqn\dw{ \eqalign{
|\phi|=0, \;\;\;\; M_{ds}^2=A>0 ~.}}
The effective potential is obtained by integrating $M_{ds}^2$,  as 
 shown qualitatively in Fig.(2).
For small $\phi$,  $V_{eff}$  can  be easily
calculated 
\eqn\dw{ \eqalign{
V_{eff}(\phi)\;
= \; \int\displaylimits_0^{\phi}\;\phi~M_{ds}^2~d\phi
&\sim \; \int\displaylimits_0^{\phi}\;\phi(A\;+\;{16{\pi^2}g\over N}
{\phi}^2)\;d\phi ~,\cr
&= \;{A\over 2}{\phi}^2\;+\;{4{\pi^2}\over {96\pi^2\;+\;
	\lambda}}{\lambda\over N}\phi^4 ~.\cr}}
For   $\phi>\phi_{max}$,
 the effective mass squared $M_{ds}^2$ turns   complex and
the effective potential has an imaginary part, which is related to
the particle creation in the scalar field $\phi$ \liN.
The imaginary $V_{eff}$
and the vacuum instability imply that one can actually deal
with the vacuum state with the field $\phi$ only during such
a short time interval, for which the energy density of the
particles created during this time,   will be much less than
Re$(V_{eff})$, and the effects of the tachyonic instability of
vacuum can be neglected.

The primordial power spectrum of the scalar field   
is obtained by taking
the long-wavelength limit of the two point correlation function,
\eqn\dw{ \eqalign{
P(k, \eta)&=\;\langle \phi_k^\alpha(\eta)\phi_{-k}^\alpha(\eta)\rangle ~,\cr
&= \;\; a(k)^2 J_\mu^2(k\eta)  ~,\cr
&\sim  \;\; {1 \over {k^{2\nu}}} ~,\cr
&= \;\;  k^{-3+ {2\over3}{M_{ds}^2\over H^2 ~}}
\buildrel \lambda \rightarrow 0 \over \longrightarrow~~ 
k^{-3+ {1\over6\pi} \sqrt{\lambda\over2}}  ~.
}}

\vskip 10pt
\centerline{\bf 4.2 Intermediate Coupling:$\;\lambda\sim N$ }
As $\lambda\;$ is increased to the order of  of $N$, 
the effective mass squared $M_{ds}^2$ turns to be a double-valued 
function of $\phi^2$, as illustrated in Fig.(3). 
A new $O(N)$-symmetric solution C emerges  with  $M_{ds}^2\sim H^2$.
Vacuum solution C has a lower  effective potential 
compared to  vacuum A. This can be shown by 
taking the integration of $M_{ds}^2$ from A to the 
transition point B and then from B to C 
\eqn\dw{ \eqalign{
V(C)\;=\; \int\displaylimits_A^B\phi M_{ds}^2(\phi)d\phi\;
	+\;\int\displaylimits_B^C \phi M_{ds}^2(\phi)d\phi
+V(A) \;<\;V(A)  ~.}}
Thus, state  C is the true ground state, as  illustrated in 
Fig.(4).
The large effective mass vacuum state 
is an unique solution in the large $N$ approximation, which is 
absent in the traditional one-loop analysis \ref\lii{A.D. Linde, Phys. Lett. {\bf 114B}, 431 (1982).}.
At the limit of ${M_{ds}^2\over H^2}\rightarrow {9\over4}$, we obtain
almost  Gaussian white noise spectrum  with $\nu\rightarrow 0$
$$
{M_{ds}^2\over H^2}={9\over 4}-\nu^2,~~~~~
m^2=-{\lambda H^2\over 90 \pi^4} {1\over \nu^2} ~.$$
The leading order of the  gap equation \gap~is
\eqn\sha{ \eqalign{
\nu^2\log(\nu)+{4(96\pi^2+\lambda)\over 87\cdot90\pi^4}=0 ~.
}}
Numerical simulations show that small $\nu$ solution 
is obtained  around $\lambda\sim N$ and the vacuum state
C exists at the range of $0<\nu<0.8$.  And  the power spectrum  
is
$$
P_k\approx {1\over k^{2\nu}}\approx {1\over k^{0\;\sim\;1.6}} ~.
$$

For very large $\lambda$, or $\lambda \gg N$, 
the contribution from high order terms will be significant and the  
the $1/N$ expansion is no longer  reliable \liN.
The power spectrum index $2\nu$ versus $\lambda$ is plotted in Fig.(5).

\vskip 10pt
\centerline{\bf 5. Verification of the Renormalization and Matching Schemes}
The divergence  of two-point function  $\langle \phi^2\rangle$ 
in de Sitter space 
can be removed  by  a less formal  but  more intuitive methods:
inflationary stage naturally provides  both  ultraviolet and 
infrared cutoffs  at momentum space.
Let's  consider
the physical wavelength of mode $k,\; \lambda_{phy}\equiv a(\eta)/k$. 
 Any mode with  physical wavelength greater than
 the Hubble radius  at the beginning of inflation ($\eta=\eta_i$), or 
\eqn\zhu{ \eqalign{
\lambda_{phys}>1/H\;\Rightarrow\; k\,<H,\;\; \;\;{\rm with}  
\;\eta_i= -{1\over H }
\;\; \;{\rm and}\;\; a(\eta_i)= 1 ~,\cr
}}
would be totally stretched 
out by inflation and is irrelevant to the physical processes after
 inflation. Also,  
any mode function with physical wavelength 
shorter than the Hubble radius  after inflation
will get more and more subhorizon in FRW, so it  does not contribute to
$\langle \phi^2\rangle$ either,
\eqn\hu{ \eqalign{
\lambda_{phys}<1/H\;\Rightarrow\; k\,>1/\eta_0\;\;   ~.
}}
The same scheme was  applied by Vilenkin \renorm~
to study the quantum 
fluctuation of a scalar field during inflation.

We are most interested in  the  small coupling region, where
 ${M^2_{ds}\over H^2} \ll 1$ and $\nu \sim {3\over2} 
-{1\over 3}{M^2_{ds}\over H^2}$.
From equation \eg
\eqn\dw{ \eqalign{
\langle \phi^2\rangle =N\int\displaylimits_{H}^{1/\eta_0}
 {d^3k \over (2\pi)^3}
\phi_k^2(k\eta_0)&= {N\pi H^2\over 4} \int\displaylimits_{H}
^{1/\eta_0}{d^3k \over (2\pi)^3}\eta_0^3 \biggl[ H^{(2)}_\nu
(k\eta_0)\biggr]^2  ~,\cr
&\sim {N\over 16}{H^2\Gamma^2(\nu) 4^{2\nu}\over \pi^2 }
\int\displaylimits_{0}^{1} d(k\eta_0) (k\eta_0)^{2-2\nu} ~,\cr 
&\sim 4N{H^2 \Gamma^2(\nu) \over  \pi^2 } {1 \over 
3-2\nu} ~. \cr 
}}
In  FRW space, the  mode functions satisfy equation \ddww
\eqn\dw{ \eqalign{
\phi_0^2 &\sim N{2H^2\nu^2 \over \pi^3} (\Gamma(\nu) \Gamma(\mu))^2
{ \Gamma(\mu-1/2)\over\Gamma(\mu+3/2)} ~,\cr
&\sim {9N\over 2}{ H^2 \Gamma^2(\nu) \over  \pi^2 } {1 \over
3-2\nu} ~. \cr}}
 The $\langle \phi^2 \rangle $ terms  match remarkably well from de Sitter stage
to FRW stage in this limit.

Around the Gaussian white noise limit, we have ${M_{ds}^2 
\over H^2}\sim 9/4,\> \;\nu\sim 0,
\;\;$ and $\> \phi_\nu(x)\sim {i\pi\over 2}ln({2\over x}) $.
The de Sitter solution has
\eqn\dy{ \eqalign{
\langle \phi^2 \rangle=N\int\displaylimits_{H}^{1/\eta_0} {d^3k \over (4\pi)^3}
\phi_k^2(k\eta_0)&\sim { N H^2\over 16 \pi^2 }\int\displaylimits_{0}
^{1}d(k\eta_0) (k\eta_0)^2 ln^2({2 \over k\eta_0})
~. }}
And the mode functions of FRW satisfy equation \ddww
\eqn\dz{ \eqalign{
\phi_0^2 &\sim N {H^2\over \pi^2}\int{ {d(k\eta)\over (k\eta)^2}ln^2({2\over\k\eta_0})
J_\mu^2(k\eta_0)} ~,\cr 
&\sim {N H^2\over 64\pi^2} \int{
d(k\eta_0) (k\eta_0)^2 ln^2({2 \over k\eta_0})} ~.
}}
Equation \dy ~and equation \dz~are dominated by small argument integrand, 
so they match each other relatively well.

The  consistency among above  two schemes  of renormalisation 
verifies the soundness of our approaches.

\centerline{\bf 6. Conclusion}
In conclusion of the paper, topological defect
originated from  quantum fluctuations during 
inflation  is a viable
and  promising picture for cosmic structure formation. 
We show that the de Sitter vacuum state  provides a
natural initial condition for the symmetry breaking 
scalar field in FRW, which is described by the classical
nonlinear sigma model. 
Due to an early stage of inflation, the scalar field 
possesses superhorizon correlations. Thus, this scenario
might  have more large scale power than 
usual cosmic defects models.  
By proper adjustment of the only free parameter: 
the coupling constant $\lambda$, 
we can obtain power spectrum ( $P_k=k^{-2 \nu}$) 
ranging  from exact scale-invariant to Gaussian white noise. 
 In the  weak coupling case ($\lambda\ll N$),  we have $2\nu= 3-\delta$,
 where $\delta\sim {1\over 6\pi}\sqrt{\lambda\over2}$. 
For intermediate coupling region ($\lambda\sim N$), $2\nu$ takes the value of 
$0$ to $1.6$.

\vskip 10pt
\centerline{\bf  7. Acknowledge}
We would like to thank M. Bucher, A.D. Linde for helpful comments and discussion.
This work was supported by NSF contract
PHY90-21984, and the David and Lucile Packard Foundation.

\vfill
\eject

\centerline{\bf Figure Captions}

\fig{1}{The effective mass squared $M_{ds}^2$ versus 
$\phi$ in the limit of weak
coupling with $\lambda\ll N$. 
At  the symmetric  phase point A: $ \phi^2=0$~ and $M_{ds}^2(A)>0$.}

\fig{2} {The effective potential $V_{eff}$ in the weak-coupling limit. 
The effective potential is calculated by integrating
 $M_{ds}^2$.  The  $O(N)$-symmetric ground state lies in $\phi^2=0$. For 
$\phi^2 >\phi_{max}^2$, the effective potential turns imaginary
representing the generation of classical field.}

\fig{3} {The effective mass  $M_{ds}^2$ versus $\phi$ in the intermediate 
interaction range with  $\lambda \sim N$. New vacuum solution C 
emerges in the limit of large effective mass. The positive
$|\phi|^2$ region is divided into two separated regions, 
from A to B, and from B to C.}

\fig{4} {The effective potential $ V_{eff}$ in the intermediate coupling region.
The $O(N)$-symmetric  ground state is the large mass solution C which
has lower effective potential.}

\fig{5} {Power spectrum index $2\nu$  with respect to $\lambda$. 
($P_k=k^{-2\nu}$)
At $\lambda\ll N$, we have the almost scale-invariant spectrum
$2\nu\sim 3$. The vacuum state jumps from A to C when $\lambda$ 
reachs the order of $N$. Gaussian white noise limit is obtained
around $\lambda\sim N$. There is no solution for $\lambda \gg N$.}

\vfill
\eject

\baselineskip=18pt

\listrefs

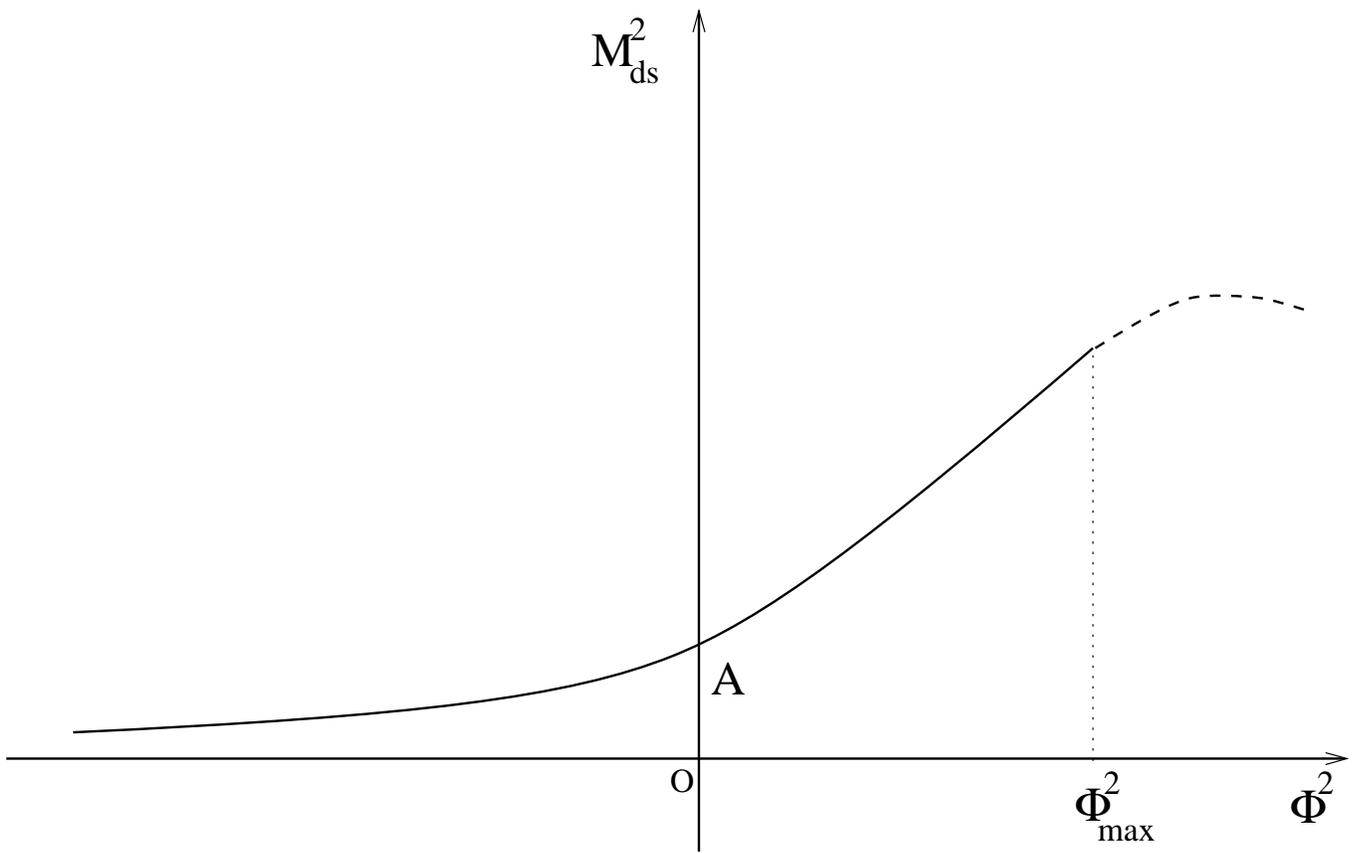
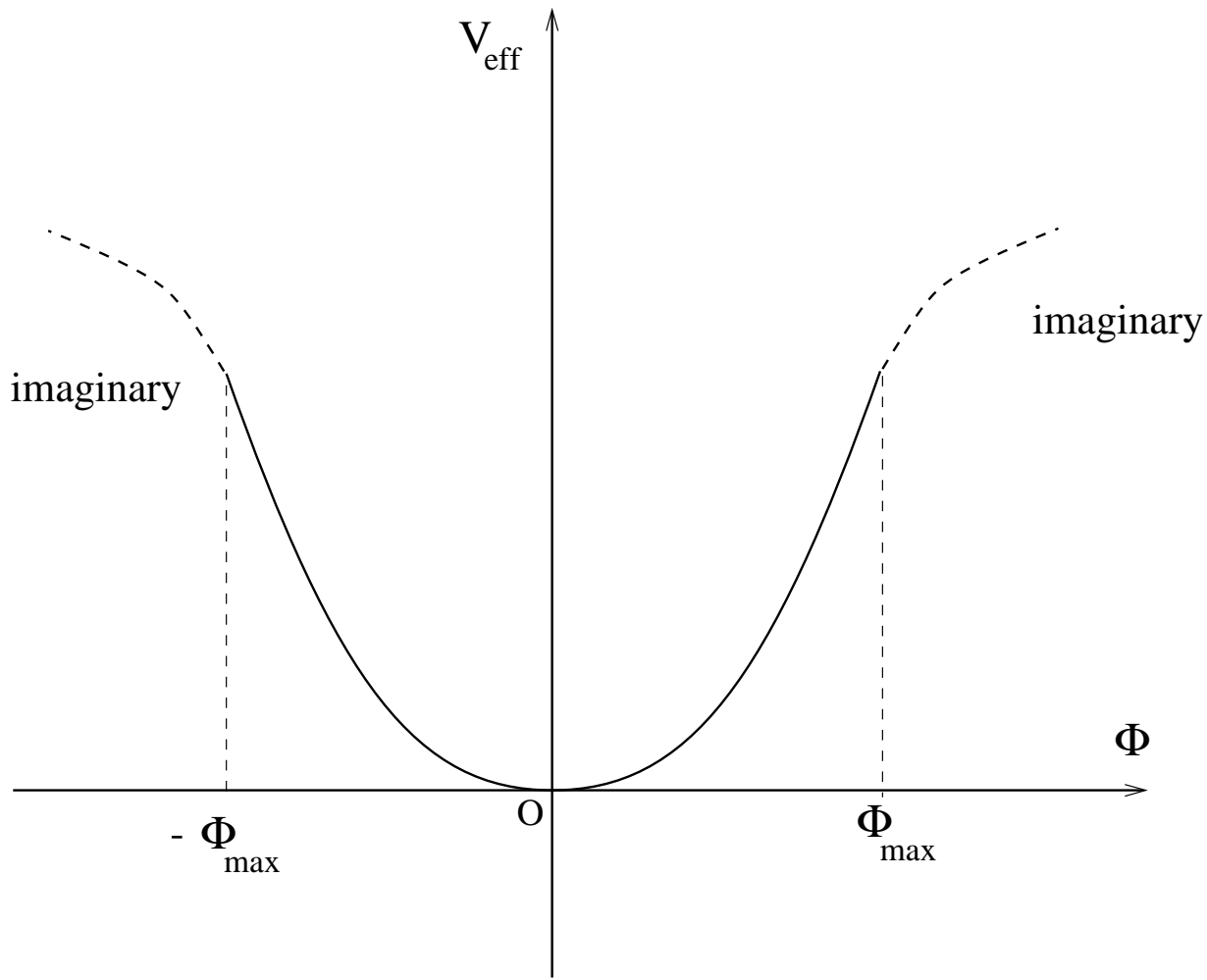
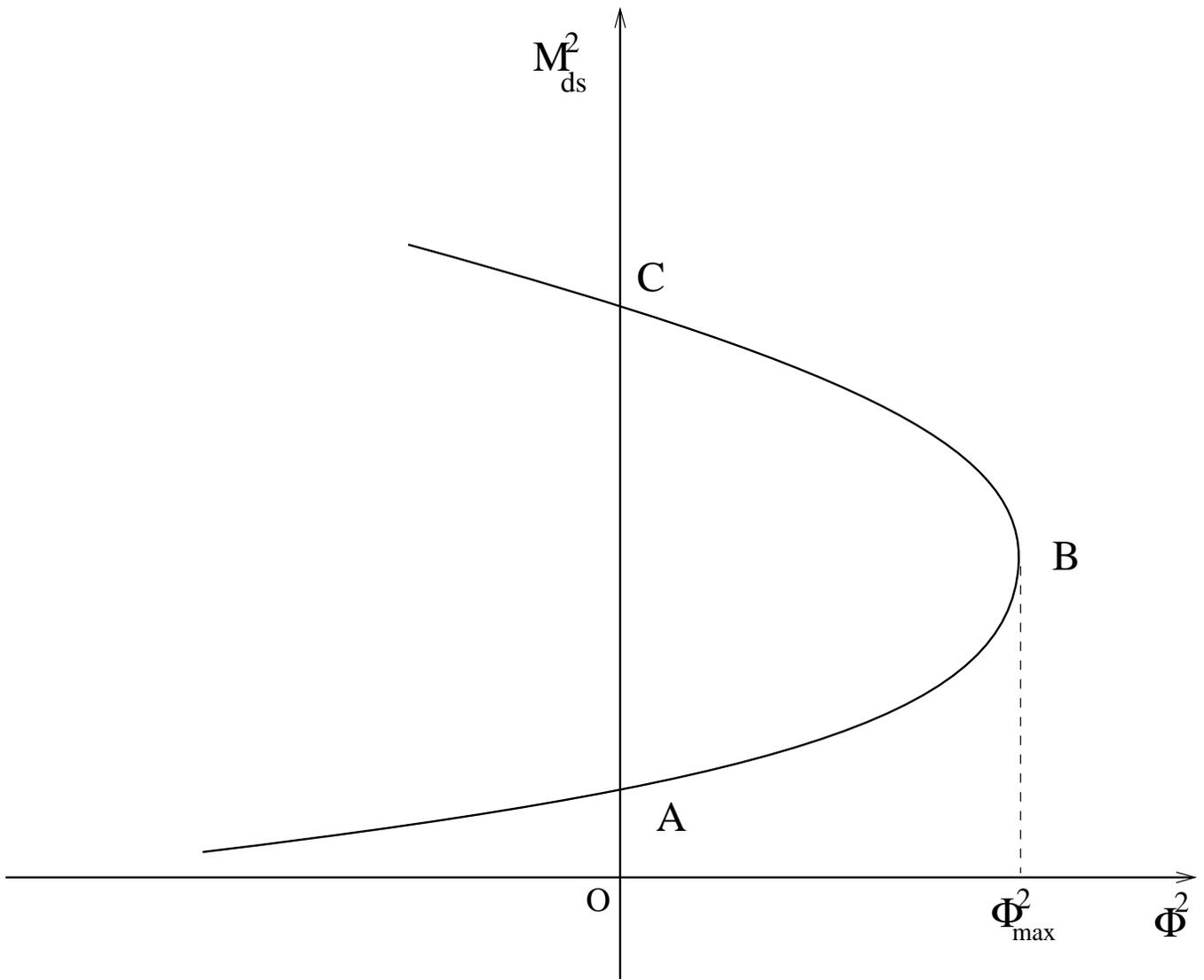
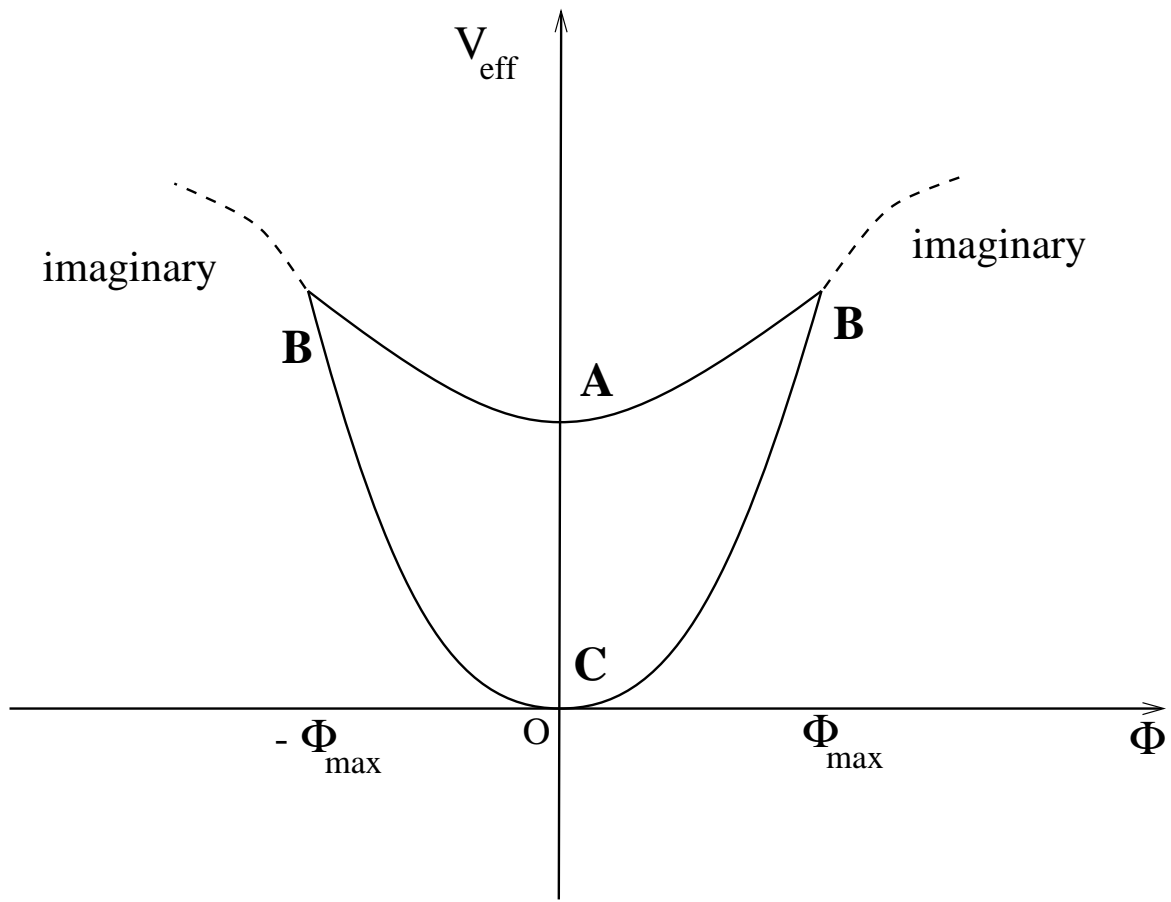
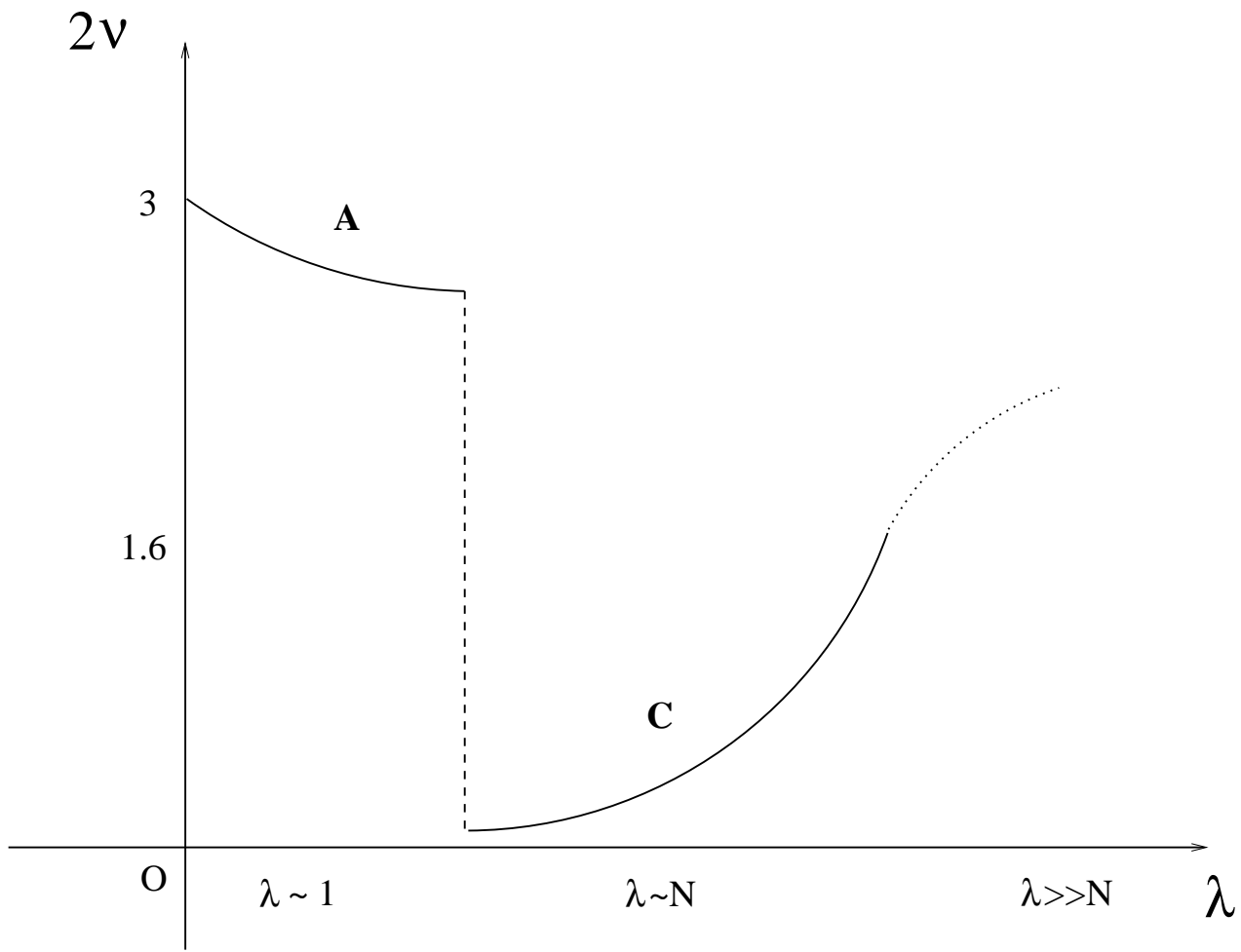
\bye